\documentclass[twocolumn,prc,amssymb, aps, showpacs,preprintnumbers,
amsmath,floatfix]{revtex4}

\usepackage{color}
\usepackage{amsmath,mathtools,booktabs,
            microtype} 

\usepackage{epstopdf}
\usepackage{graphics}
\usepackage{graphicx}
\usepackage{dcolumn}
\usepackage{bm}
\usepackage{epsfig}
\usepackage{times}
\usepackage{url}

\begin{document}

\title{Coupling highly excited nuclei to the atomic shell in dense astrophysical plasmas}

\author{Stephan \surname{Helmrich}\footnote{Present address: Physikalisches Institut, Universit\"at Heidelberg, Im Neuenheimer Feld 226, 
D-69120 Heidelberg, Germany}}
\email{helmrich@physi.uni-heidelberg.de}

\author{Katja \surname{Spenneberg}}

\author{Adriana \surname{P\'alffy}}
\email{palffy@mpi-hd.mpg.de}

\affiliation{Max-Planck-Institut f\"ur Kernphysik, Saupfercheckweg 1, D-69117 Heidelberg, Germany}
\date{\today}

\begin{abstract}
In dense astrophysical plasmas, neutron capture populates highly excited nuclear states close 
to the neutron threshold. The impact of additional low-energy nuclear excitations via coupling 
to the atomic shell on the ability of the so-formed compound nucleus to retain the captured neutron is investigated. 
We focus on the mechanism of nuclear excitation by electron capture in plasmas characterized by electron fluxes 
typical for the slow neutron capture process of stellar nucleosynthesis. The small effect of this further excitation on the neutron capture and gamma decay sequence relevant for nucleosynthesis is quantified and  compared to 
the corresponding effect of an additional low-energy photoexcitation step.

\end{abstract}


\maketitle

\section{Introduction}
At the interface between nuclear and atomic physics, a special role is played by nuclear processes that directly involve atomic electrons. For instance, it is well known that the population and lifetime of nuclear excited states can be affected by the electronic shells in the processes of nuclear electron capture (EC) and internal conversion (IC). Counterintuitive examples where more electrons available for IC or EC do not necessarily lead to a shorter nuclear excited state lifetime have been experimentally observed, for instance in $^{57}_{26}\mathrm{Fe}$ where decay measurements  of the 14.4~keV M\"ossbauer level  in one- and two-electron ions have shown that the nuclear lifetime is about $20\%$ shorter in H-like $\mathrm{Fe}^{25+}$ ions   than  in He-like  $\mathrm{Fe}^{24+}$ ions or in the neutral atom \cite{Philips}.  Similarly,  experimental results have been obtained for EC rates in H-, He-like and neutral $^{140}_{59}\mathrm{Pr}$, where the nuclear lifetime of the one-electron $\mathrm{Pr}^{58+}$ ion is shorter than the ones of the corresponding two- or many-electron cases \cite{Yuri}. The inverse processes of IC and EC, namely, nuclear excitation by electron capture (NEEC) and bound $\beta$ decay, respectively, require the presence of vacancies in the atomic shell. NEEC in highly charged ions followed by x-ray emission has been shown to  prolong by two orders of magnitude the lifetime of excited states in actinides \cite{Pa08}. More spectacularly, the opening of the new bound $\beta$ decay channel in highly charged ions  influences the half life of unstable levels in nuclei \cite{Takahashi1983,Takahashi2} and via this mechanism the   ground state $^{187}_{75}\mathrm{Re}$ lifetime decreases by more than nine orders of magnitude  from 42~Gyr for the neutral atom to 32.9 yr for bare ions as a consequence of new bound $\beta$ decay branches to the ground and excited states of the $^{187}_{76}\mathrm{Os}$ daughter \cite{Bosch2,bound_beta_clock}. The case  of $^{187}_{75}\mathrm{Re}$ is particularly interesting in astrophysical context, since  it concerns the accuracy of the $^{187}\mathrm{Re}-^{187}\mathrm{Os}$ cosmochronometer \cite{bound_beta_clock}. The  behavior of nuclei in highly charged ions is thus of potential interests in nuclear astrophysics and  studies of nuclear decay properties. 

\begin{figure}[!h]
\vspace{-0.2cm}
\centering
\scalebox{0.5}{\includegraphics{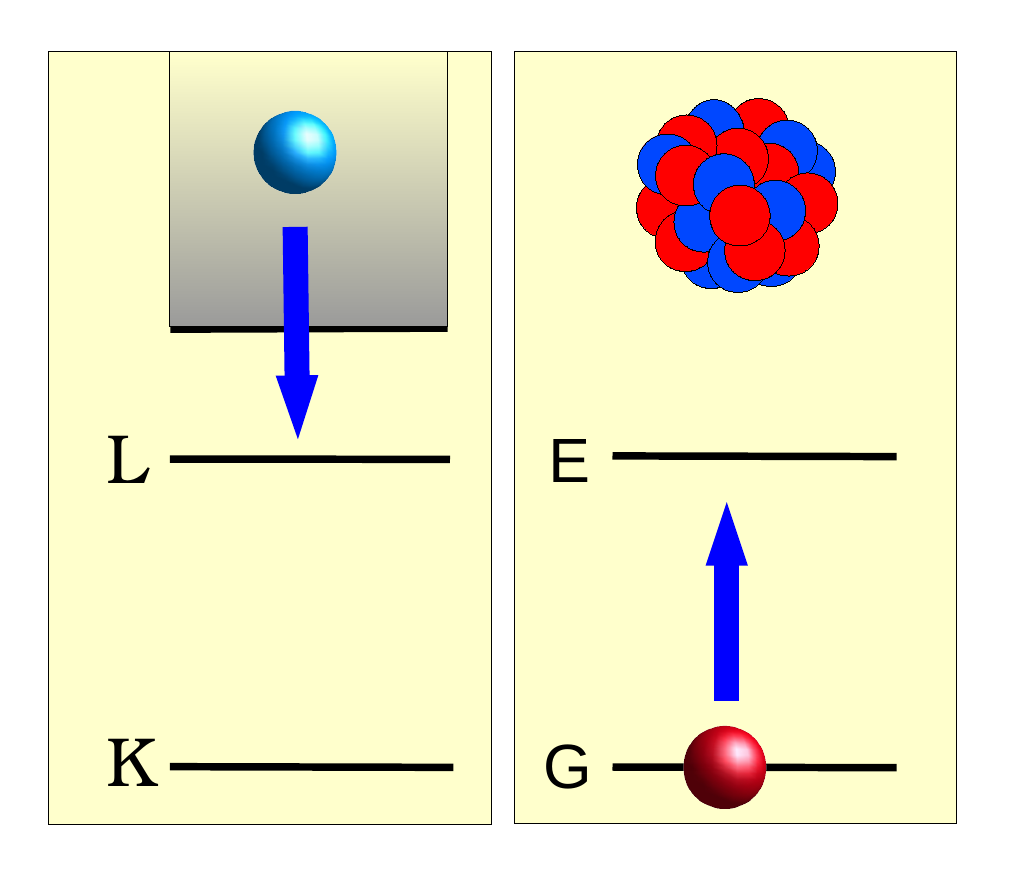}}
  \caption{\label{fig1} (color online). Schematic illustration of NEEC. If the atomic and nuclear transition energies match, an electron can recombine into an ion (depicted in the left panel by the bare $K$ and $L$ atomic shells) with the simultaneous excitation of the nucleus (right panel) from the ground state $\mathrm{G}$ to the excited state $\mathrm{E}$.}
\end{figure}

So far, the role of NEEC was never a subject of sustained investigation in nuclear astrophysics. In the resonant process of NEEC, a free electron with matching kinetic energy recombines into  a highly charged ion with the simultaneous excitation of the nucleus, as it is schematically shown in Fig.~\ref{fig1}. As nuclear excitation mechanism, NEEC becomes increasingly efficient with rising electron density and degree of ionization. These conditions are predominant in dense astrophysical plasmas in the interior of stars and supernovae. In the context of isomer depletion, NEEC and its sibling nuclear excitation by electron transition (NEET) populating low-lying nuclear excited states under dense plasma conditions have been investigated \cite{Gosselin04, Gosselin07, Gosselin10, Morel2010,Gunst2014}, predicting  an enhancement of the isomer state decay up to several orders of magnitude. As resonant electron recombination channel, NEEC favors free electrons with low kinetic energy.  Fast electrons are less likely to recombine such that the amount of energy that can be transferred to the nucleus is limited; when starting from the ground state, typically only nuclear excitation  to low-lying excited states occurs. 
In nucleosynthesis models, the sometimes significant excitation of such levels due to the interaction with the hot thermal photon bath is accounted for by assuming thermally  equilibrated nuclei \cite{s_process_review}.  This procedure does not address explicitly the particular transition mechanisms, and  all the information on the thermal population of low excited states is captured in the stellar enhancement factor (SEF) \cite{bao:neutron, Rauscher2000}.

In this work we investigate  a different scenario, in which NEEC occurs not from the nuclear ground state or metastable states in its vicinity, but instead from  highly excited states close to the neutron threshold. An intrinsic assumption of neutron capture nucleosynthesis models is that the formed compound nucleus decays practically instantaneously from an initial state with $E\simeq S_n+kT$, where $S_n$ denotes the neutron separation energy, to a thermal distribution of low-lying nuclear levels \cite{B2FH,s_process_review}. Only recently it was argued in Refs.~\cite{Bernstein,Bernstein2} that the  neutron capture may be followed by the absorption of a low-energy photon (less than $1\,\mathrm{MeV}$) prior to  statistical $\gamma$-ray emission. This effect was found to be temperature dependent and to dominate over the direct $\gamma$ decay above a certain temperature typical for rapid neutron capture nucleosynthesis (r-process). Since the decay branching ratio of the compound state is highly sensitive to its energy, even excitation by a small energy amount would lead to an effective reduction of the neutron separation energy as a function of the environment temperature and density and would eventually  shift the nucleosynthesis path towards the valley of stability. The purpose of the present work is to find out whether such an effect can also be expected when considering NEEC  as excitation mechanism instead of photoabsorption.   Resonant electron recombination mechanisms like NEEC are believed to be the dominating form of recombination in hot astrophysical plasmas \cite{Massey1942}, where high degrees of ionization and high electron densities prevail. Additionally, theoretical values show that excitation of the nucleus by coupling to the atomic shell can be  more efficient than photoabsorption for nuclear transitions of low energy \cite{Pa07,PaJMO}. An estimate of the possible impact of NEEC occurring on compound nuclei formed by the capture of a slow neutron would be a useful counterpart of the results on r-process in Refs.~\cite{Bernstein,Bernstein2}. We therefore investigate possible changes via the additional nuclear excitation on the neutron capture and $\gamma$ decay sequence relevant for s-process nucleosynthesis.

The NEEC formalism developed in Ref. \cite{Pa06} for nuclear transitions close to the ground state is adapted to describe excitation starting from the compound state.   For  nuclear excitation energies  on the order of several $\mathrm{MeV}$, nuclear states are rather described by level densities than discrete spectra. The nuclear matrix elements in the NEEC transition rates, for transitions between discrete levels given by the reduced nuclear transition probabilities, are here estimated with the help of a nuclear level density parametrization obtained from the photon strength function of the giant dipole resonance~\cite{Greiner}. Furthermore, we extend the theoretical NEEC treatment to take into account the relevant temperature domain of both neutron and electron fluxes as well as the role of multiple neutron resonances of the neutron capture spectrum. The impact of NEEC is quantified by defining a stellar mitigation factor as the counterpart of SEF taking into account the additional excitation of the compound nucleus. Our results for the numerical examples $^{187}\mathrm{Os}$ and $^{193}\mathrm{Ir}$ show that for typical s-process nucleosynthesis conditions, the effect of  coupling highly excited nuclei to the atomic shell is small, with SMF values on the  order of $10^{-9}$ raising to $10^{-4}$ only for  a plasma temperature of $T\simeq 1.2\cdot 10^9$ K. A comparison with the results in Ref. \cite{Bernstein} shows that such higher plasma temperatures as typical for r-process nucleosynthesis equally not favor the coupling to the atomic shells, but rather photoabsorption.

The paper is organized as follows. In Sec.~\ref{scheme} we present the model used to include the additional excitation of the compound nucleus in astrophysical scenarios. This is followed by an outline of the NEEC rate calculations in dense plasmas considering multiple neutron resonances. Our numerical results are presented in Sec.~\ref{results}.  The paper concludes with a Summary in Sec.~\ref{conclusion}.

\section{Theoretical framework for NEEC at the neutron threshold \label{scheme}}
\begin{figure}[h]
\begin{center}
  \includegraphics[width=\linewidth]{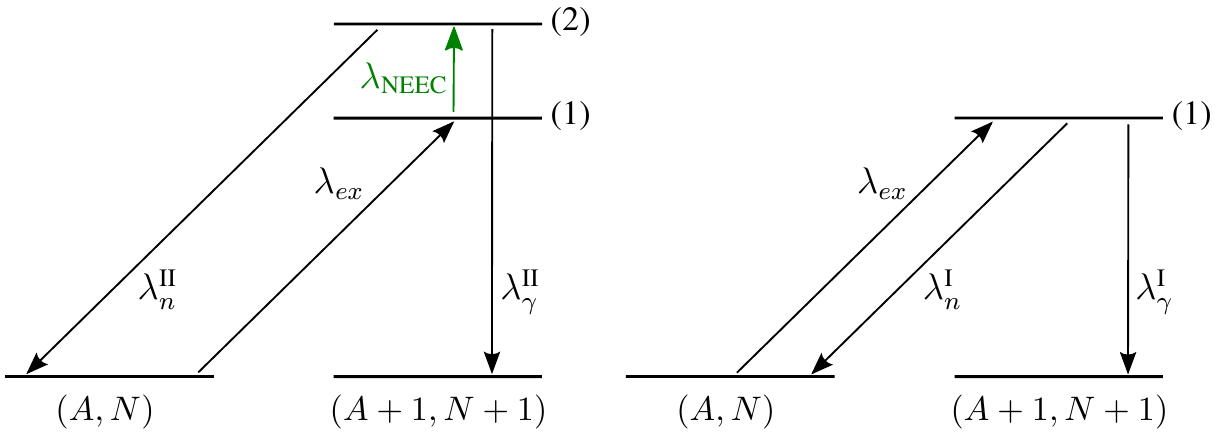}
  \caption{Standard neutron capture process (right) and sequence including the additional NEEC excitation (left). Due to the higher excitation energy after NEEC, the decay properties of state (2) may significantly differ from the ones of state (1).}
  \label{im:fig2}
\end{center}
\end{figure}

The slow neutron capture process (s-process) of nucleosynthesis comprises a repeated sequence of neutron capture, $\gamma$ decay of the excited compound state and successive $\beta$-decay events. The time scale of the s-process is  dominated by the (slow) neutron capture. It is therefore justified to restrict our analysis of the impact of NEEC on s-process nucleosynthesis to the excitation and decay steps prior to $\beta$ decay. We  address  a  sequence of processes encompassing  the initial neutron capture event, NEEC and the $\gamma$-ray cascades and neutron re-emission. The nuclei of interest are initially excited to a highly energetic compound state by neutron capture. This state may decay right away via $\gamma$ decay or neutron re-emission. For the initial capture of slow neutrons, $\gamma$  decay is typically more probable than neutron re-emission.  However, the short finite lifetime of the compound state may allow processes like NEEC or photoabsorption  to occur in the plasma electron- and photon-bath prior to  deexcitation. These processes further add energy to the compound nucleus, and even few keV may change the decay properties of the compound nucleus which then consequently favors neutron re-emission over $\gamma$ decay. The  scenario we consider to quantify the impact of NEEC on nucleosynthesis has the following steps, schematically illustrated  in Fig.~\ref{im:fig2}: 
\begin{itemize}
	\item Formation of the compound nucleus by neutron capture with the excitation rate $\lambda_{ex}$.
	\item The possible direct decay by neutron and photon emission of the compound nucleus with the associated rates $\lambda_\gamma^\text{I}$ and $\lambda_n^\text{I}$. At neutron resonance typically $\lambda_\gamma^\text{I} / \lambda_n^\text{I} \sim 10$. 
	\item Alternatively the consecutive excitation of the compound nucleus by NEEC to a higher excited state with $\lambda_\text{NEEC}$.
	\item The final decay of the second excited state via neutron emission $\lambda_{n}^\text{II}$ or gamma decay $\lambda_{\gamma}^\text{II}$ with $\lambda_{\gamma}^\text{II} / \lambda_{n}^\text{II} \sim 0.1$. 
\end{itemize}

At excitation energies close to the neutron threshold discussed in this scenario, the nuclear spectrum  enters a continuum regime described by  level densities which increase with excitation energy. The probability of a NEEC step built on the nuclear compound state is therefore larger compared to the corresponding IC decay mediated by the couping to the atomic shell. Our approach is to consider the two-step process of NEEC and the subsequent nuclear decay.  The  total cross section for this two-step process can be derived in a general manner. Cross sections and flux of continuum particles combine to a reaction rate of NEEC (induced by the capture of electrons into the atomic shell) and subsequent decay which can be compared to direct---and thus electron-flux independent---decay rates. 

We quantify the implications of additional NEEC or photoabsorption on the s-process nucleosynthesis scenario defining, in analogy to the SEF, a stellar mitigation factor (SMF) as the ratio of the effective cross sections  for the  scenarios taking into account (II) and without accounting for (I) NEEC,
\begin{equation}
\label{eq:smf2}
  \text{SMF} = \frac{\langle \sigma v \rangle^\text{II}}{\langle \sigma v \rangle^\text{I}}\, . 
\end{equation}
In the equation above, the product of the cross section $\sigma$ and the velocity of the incoming neutrons $v$  are averaged over to render the reaction rate of the considered sequence of processes. As a relevant quantity, we will investigate the SMF for the cross section $\sigma_{(n,\gamma)}$ for neutron capture followed by $\gamma$ decay. In the following subsections we address and deduce step by step all the quantities required for the SMF calculation.

\subsection{Maxwellian-averaged cross sections}\label{sec:rates_general}
In order to derive the stellar reaction rate $\lambda = \langle \sigma v\rangle$ from the cross section one has to integrate over the velocity distribution $\varphi(v)$ of incoming particles  to obtain 
\begin{equation}
\label{lambdaX}
 \lambda = \langle \sigma v\rangle = \int \sigma v \varphi(v)\, \mathrm{d}v\,. 
\end{equation}
Writing the above formula  in terms of energy renders
\begin{equation}\label{reactionRateGeneral}
 \lambda = \int \sigma(E) v(E) \varphi(E)\, \mathrm{d}E = \int \sigma(E) \phi(E)\, \mathrm{d}E\,  ,
\end{equation}
with $\phi(E) \mathrm{d}E$ the relativistic differential flux. Impinging particles of interest in our model are neutrons and electrons, both being fermions. The differential flux for one species of incoming particles is hence given by 
\begin{eqnarray}\label{flux}
 \phi(E)\, \mathrm{d}E & =& v(E) \varphi(E)\, \mathrm{d}E \nonumber \\
  & =& \frac{\mathrm{d}E}{\pi^2 \hbar^3 c^2}  \frac{(E^2 + 2Em c^2)}{1+\exp[(E-\mu)/k_B T]}\,  ,
\end{eqnarray}
with $\mu$ the chemical potential, $c$ the speed of light, $k_B$ Boltzmann's constant, $m$ the particle (electron, neutron) mass and $T$ the plasma temperature, respectively. The chemical potential can be set by choosing as normalization of the energy-distribution function $\varphi(E)$ the number density $n$ of the respective free particles in the plasma,
\begin{equation}\label{number_density}
 n = \frac{N}{V} = \int_0^\infty \varphi(E) \, \mathrm{d}E\,. 
\end{equation}
The numerical value of the number density $n$  can be derived from experimental data, and the equation above  can then be numerically solved for $\mu$. In the following we address the remaining term that determines the rate $\lambda$, namely the cross section for the considered sequence of processes.

\subsection{Cross section expressions}
The general cross section for resonant excitation via capture of continuum particles with energy $E$ is given by 
\begin{equation} \label{cross_Lo}
 \sigma_{ex}(E) = \frac{\pi \hbar^2}{p^2} \frac{\Gamma_{ex} \Gamma_{tot}}{(E - E_{res})^2 + (\Gamma_{tot}/2)^2}\, , 
\end{equation}
where $\Gamma_{ex}$ is the width corresponding to the excitation rate and $\Gamma_{tot}$ the total width of the resonance. Furthermore, $p$ denotes the momentum of the incoming particle and $E_{res}$ the position of the resonance.   Subsequent decay into a channel $\delta$  is incorporated by a branching ratio 
\begin{equation} \label{sig_general}
	\sigma_{(ex, \delta)}(E) = \sigma_{ex}(E)  \frac{\Gamma_\delta}{\Gamma_{tot}}\, ,
\end{equation}
where $\sigma_{tot} = \sigma_{ex} = \sum_\delta \sigma(ex, \delta)$ \cite{Blatt_weisskopf}.

The general expression in the equations above forms the basis of many stellar reaction model calculations,  for instance of the Maxwellian-averaged capture cross sections (cf. Refs.~\cite{n_tof_III, Rauscher2000}). 
In the following we proceed to formulate the cross section expression for NEEC followed the nuclear decay via the decay channel $\delta$. We assume that the nucleus is initially in the compound state with energy $E_{\alpha}$ to which it was excited by neutron capture. The compound state energy is given by the neutron separation energy $S_\text{n}$ and the additional  neutron resonance energy $E_{n}^\text{res}$ corresponding to the  kinetic energy of the captured neutron, 
\begin{equation}
 E_{\alpha} = S_\text{n} + E_{n}^\text{res}\, . 
\end{equation}
The nuclear energy is further increased starting from $E_{\alpha}$ via NEEC on the compound state causing changes of the nuclear decay channels.

The energy transferred to the nucleus by NEEC is composed of the kinetic energy of the continuum electron, henceforth denoted by $E_e$, and the (negative) binding energy $E_b$ of the electron. The latter  is given by the energy of the orbital into which the electron recombines. The desired cross section for one electron resonance thus reads 
\begin{align}\label{sig_NEEC_single}
	\sigma_{(\text{NEEC},\delta)}(E_e) = \frac{\pi \hbar^2}{p_e^2}  \Gamma_{\text{NEEC}}(E_\alpha \rightarrow E_\alpha+E_e-E_b) \nonumber \\ \times \frac{\sigma_\delta(E_\alpha+E_e-E_b)}{\sigma_{tot}(E_\alpha+E_e-E_b)}  \frac{\Gamma_{tot}}{(E_e - E_e^\text{res})^2 + (\Gamma_{tot}/2)^2} \, .
\end{align}
Here $E_e^\text{res}$ is the continuum electron energy corresponding to the resonant NEEC capture and $\Gamma_{tot}$ the total width of the resonant state, in our case the state (2) in Fig.~\ref{im:fig2}. Furthermore, $ \Gamma_{\text{NEEC}}(E_\alpha \rightarrow E_\alpha+E_e-E_b)$ is the NEEC rate starting from the compound nucleus state of energy $E_\alpha$.
The quasi continuous nuclear level spectrum reached for the compound nucleus causes the NEEC resonances to merge into a continuous spectrum which is luckily exhausted by the broad energy distribution of the continuum electrons in the plasma. The continuum NEEC spectrum is taken into account by integrating the single resonance cross section \eqref{sig_NEEC_single} over all resonances with the nuclear level density $ \rho$ as integration measure. As a simplification, we consider a generic constant value for the width $\Gamma_{tot}$ of the excited state (2) of 100 eV, consistent with the scenario discussed in Sec.~\ref{results} of NEEC into higher free electronic shells and a single $1s$ electron in the $K$ shell. In this case, the total width of state (2) is determined by the width of the electronic capture shell. The cross section for NEEC followed by decay in channel $\delta$ then reads
\begin{eqnarray}\label{eq:signa_final}
	\sigma_{(\text{NEEC},\delta)}(E_e) = \int \mathrm{d}E_e^\text{ res}\frac{2 \pi^2 \hbar^2}{p_e^2}\Gamma_{tot}\, \rho(E_{\alpha} + E^\text{res}_e-E_b)
 \nonumber \\
\times  \frac{\sigma_\delta(E_{\alpha}+E_e-E_b)}{\sigma_{tot}(E_{\alpha}+E_e-E_b)} \frac{ \Gamma_{\text{NEEC}}(E_{\alpha} \rightarrow E_{\alpha}+E_e-E_b)}{(E_e - E_e^\text{res})^2 + (\Gamma_{tot}/2)^2} \, .
\end{eqnarray}
The decay mode $\delta$ here can be either $\gamma$ or neutron emission. The corresponding nuclear cross section values can be taken from experimental data compilations. Our simulations show that there is little change in the values of $\sigma_{(\text{NEEC},\delta)}$ as a function of the total width 
$\Gamma_{tot}$. For instance, the results considering a Lorentzian resonance with 100 eV width or a delta function-like resonance differ only by a relative factor $10^{-2}$.

The NEEC width $\Gamma_{\text{NEEC}}$ can be calculated according to the formalism presented in Ref. \cite{Pa06}, which separates the electronic and nuclear degrees of freedom in the total matrix element. The nuclear part, for the case of low-lying transitions specified by the experimental reduced transition probability $B$, needs in the present case to be approximated theoretically. An approximation for the strength of the nuclear transition as well as expressions for the nuclear level densities are introduced in the next subsection.

\subsection{Reduced transition probabilities and level densities for highly excited nuclear states}\label{sec:Bees}
\begin{figure}
\includegraphics[width=0.8\linewidth]{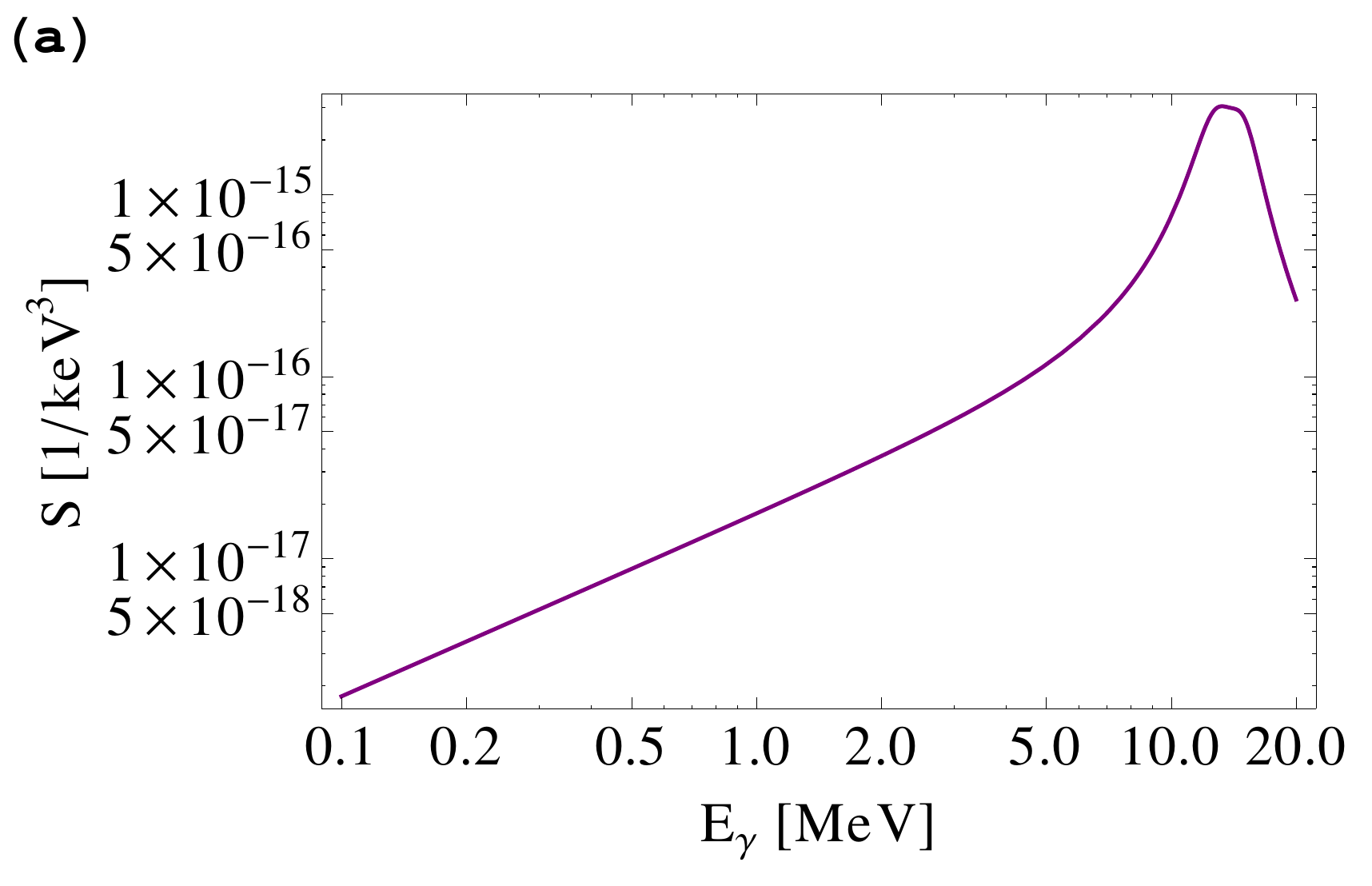}
\includegraphics[width=0.8\linewidth]{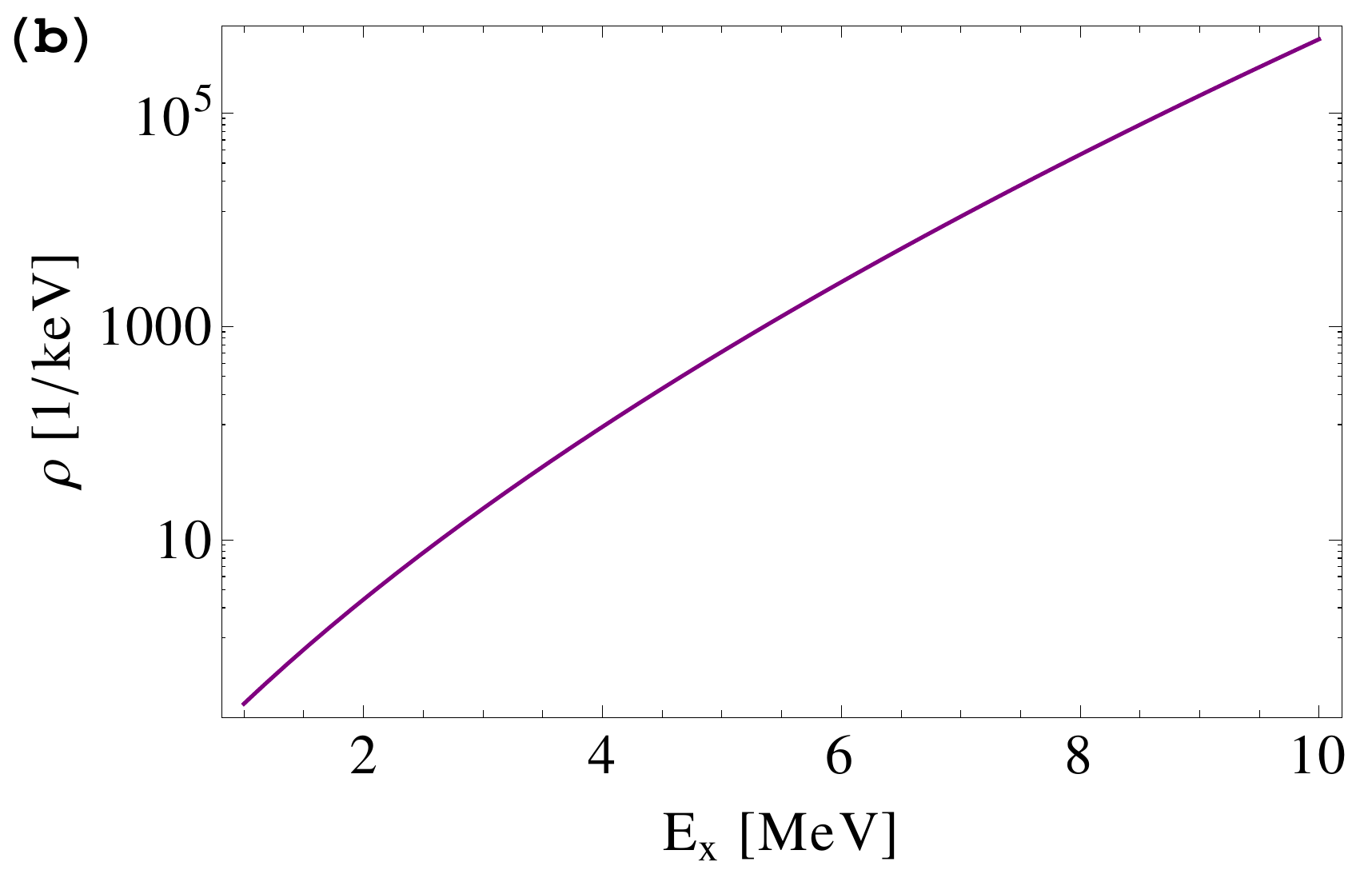}
\caption{\label{psf} (a) Photon strength function parametrization and (b)  back-shifted Fermi gas approximation of the nuclear level density  for $^{187}\mathrm{Os}$.}
\end{figure}

For nuclei, the predominant mode of photon absorption at most energies is the Giant Dipole Resonance (GDR) \cite{Greiner} typically with a Lorentzian energy dependence centered at the energy $E=80A^{-1/3}$ MeV. For axially symmetric deformed nuclei, the GDR photon strength function can  be often well fitted by a double-Lorentzian shape \cite{Becvar_psf, Becvar_psf_test},
\begin{equation}\label{eq:strength}
	S_{E1}(E_\gamma) = \frac{1}{3\pi^2(\hbar c)^2} \sum_{i = 1}^2 \frac{\sigma_i E_\gamma{\Gamma_{Gi}}^2}{\left({E_\gamma}^2 - {E_{Gi}}^2\right)^2 + {E_\gamma}^2{E_{Gi}}^2}\, , 
\end{equation}
where $E_{Gi}$ is the central energy of the axial branch $i$ of the GDR and $\Gamma_{Gi}$ its damping width \cite{Becvar_psf, psf_mod}. The transition energy is here denoted by $E_\gamma$. Parameterizations from experimental data are typically available, see Ref.~\cite{Plujko2011567}. An illustration of the parametrization for the photon strength function for $^{187}\mathrm{Os}$ is presented in Fig.~\ref{psf}(a). Here, the two Lorentzian peaks are close together and very narrow.

In 1955 Brink made the hypothesis that a GDR can be built on every excited nuclear state, not just the ground state, retaining its  shape  \cite{Brink}. For our purpose of describing a low-energy excitation of a highly excited compound state, this implies an extrapolation of the GDR parametrization down to energies on the order of $100\, \mathrm{keV}$. This assumption has been also used for calculating the impact of photoabsorption on highly excited states in r-process nucleosynthesis \cite{Bernstein,Bernstein2}, using the parametrization for low energies ($E<5$ MeV) introduced by Kadmenskij, Markushev and Furman \cite{KMF}.
Independent of the chosen parametrization, we would like to stress that using GDR strength functions for low-energy excitations  is expected to be a very rough estimate, which can be off by as much as an order of magnitude. A discussion of this high degree of uncertainty is presented in Sec.~\ref{results}.

According to Brink's hypothesis, the photon strength function $S$ of the electric giant dipole resonance relates to the expectation value of the $\gamma$-decay transition rate as 
\begin{equation}
 \overline{\Gamma}(E_x + E_\gamma \rightarrow E_x) = \frac{{E_\gamma}^3 S(E_\gamma)}{\rho(E_x + E_\gamma) \hbar}\, . 
\end{equation}
Here, the energy of the lower involved level is represented by $E_x$ and the one of the upper level as $(E_x + E_\gamma)$. Furthermore $\rho(E_x)$ is the nuclear level density associated to some small energy range $(E_x,E_x+\Delta E)$. The strength function is independent of the levels between which the transition occurs as a consequence of Brink's hypothesis. Applying the principle of detailed balance
one can derive the appropriate relationship for the inverse, upwards transition,
\begin{eqnarray}\label{photon_strength_rate}
 \frac{\overline{\Gamma}(E_x \rightarrow E_x + E_\gamma)}{\rho(E_x + E_\gamma) \; \Delta E}  &=& \frac{\overline{\Gamma}(E_x + E_\gamma \rightarrow E_x)}{\rho(E_x) \; \Delta E}  \nonumber \\
 \Rightarrow \hspace{0.4cm} \overline{\Gamma}(E_x \rightarrow E_x + E_\gamma) &= &\frac{{E_\gamma}^3 S(E_\gamma)}{\rho(E_x) \hbar}\, .
\end{eqnarray}

A general expression for the nuclear level densities as a function of energy $E$, angular momentum $J$ and parity is given by \cite{Hans,PLBHans}
\begin{equation}
\rho(E, J, \pi) = \frac{1}{2} \rho(E) \ \frac{2 J + 1}{2 \sqrt{2
\pi} \ \chi^3}  
\exp \left[ - \frac{[J + (1/2)]^2}{2
\chi^2} \right] \ .
\end{equation}
Here, the first factor $1/2$ is the parity ratio and indicates equal densities for states of either parity. The last two terms of the
product give the spin dependence, with $\chi$ the spin--cutoff
factor. With spin and parity being accounted for, $\rho(E)$ as parameterized above is
defined as the level density of spinless non--interacting Fermions
that carry no angular momentum. From the variety of models and  parametrizations describing the density of states $\rho(E)$ we employed the back-shifted Fermi gas model in the spin-independent form taken from \cite{bsfg_mod}
\begin{equation}\label{eq:rho_from_bsfg}
	\rho(E_x) = \frac{\exp[2 \sqrt{a(E_x-\Delta)}]}{12 \sqrt{2} \chi a^{1/4} (E_x - \Delta)^{5/4}}\, , 
\end{equation}
with $a$ being the single-particle density parameter, $\Delta$ the backshift, $t$ the thermodynamic temperature determined from the equation $E_x-\Delta=at^2-t$ and $\chi$ the spin-cutoff factor, given by  
\begin{equation}
	\chi^2 = 0.0146 A^{5/3} \frac{1+\sqrt{1+ta(E_x - \Delta)}}{2a}\, . 
\end{equation}
The associated fitted parameters for a great number of isotopes can be found in Ref.~\cite{bsfg_mod}. For $^{187}\mathrm{Os}$ the curve of $\rho(E)$ is depicted in Fig.~\ref{psf}(b), underlining that above neutron separation energy $S_n$ [e.g. $S_n(^{187}\mathrm{Os})=6.3\, \mathrm{MeV}$] indeed a level density approximation is appropriate. 

In Ref.~\cite{bsfg_mod}, parity asymmetry effects are explicitly disregarded, although some parity dependence of $\rho$ is known to exist especially for low-excitation energies \cite{parityasym}. How justified is such a neglect for the case under consideration here? A parametrization for the parity ratio for lower masses $20\le A \le 110$ is provided in Ref.~\cite{parityasym}.  An inspection of the parity ratio dependence on energy and mass number $A$ shows that 
at 7 MeV, the typical nucleon binding energy near the stability line, the parity ratio is close to asymptotic (i.e., $1/2$) for $A>50$. For lower binding energies, for instance 4 MeV, the parity ratio is not asymptotic until $A$ exceeds 75. 
However, regardless of the mass number $A$ and of the corresponding neutron separation energy, the
 NEEC cross section is proportional to the ratio $\rho(E_\alpha + E_e-E_b)/\rho(E_\alpha)$.
Since for all cases $E_\alpha \gg (E_e-E_b)$,  the parity asymmetry  is for all practical purposes constant on the interval $(E_\alpha, E_\alpha + E_e-E_b)$. Thus, the level density contribution for $\sigma_{\text{NEEC}}$ is given solely by the energy dependence of $\rho$ and the neglect of parity asymmetry effects is justified.

The rate defined by Eq.~\eqref{photon_strength_rate} is the average photon deexcitation rate in the considered energy regime. This can be set equal to the $\gamma$ decay rate expression used for nuclear transitions \cite{ring_schuck}
\begin{equation}\label{gamma_rate}
 A^{d\rightarrow f}(\zeta, L) = \frac{8 \pi (L + 1)}{\hbar L((2L+1)!!)^2} \left(\frac{E_\gamma}{c \hbar} \right)^{2L+1} \frac{B^{d\rightarrow f}(\zeta , L)}{4 \pi \varepsilon_0} \, ,
\end{equation}
where $\varepsilon_0$ is the vacuum permittivity, $\zeta$ the transition type [electric ($el$) or magnetic ($magn$)] and $L$ its multipolarity, respectively. 
We consider here  electric dipole transitions such that $L = 1$. The nuclear transition rate $A_r^{d\rightarrow f}(el,1)$ is equated to~\eqref{photon_strength_rate} to yield the desired nuclear transition probability. In our context $E_x = E_\alpha$ and $E_\gamma = E_e-E_b$. Hence our estimate for the nuclear transition probability reads 
\begin{equation}\label{eq:B_sum_rule}
 B(E_\alpha \rightarrow E_\alpha + E_e-E_b) = \frac{9 c^3 \hbar^3 \varepsilon_0}{4}  \frac{S(E_e-E_b)}{\rho(E_\alpha)} \, .
\end{equation}
Using this expression for the adapted reduced nuclear transition probability, the remaining electronic matrix element and finally the NEEC rate can be calculated following the formalism of Ref. \cite{Pa06}. We quote here only the final result of the NEEC rate expression for electric dipole transitions from an initial state with angular momentum $I_i$ to the final state characterized by $I_d$,
\begin{eqnarray}
Y_n & = &\frac{4 \pi^2 \rho_i^2}{9} B(E1,I_i \rightarrow I_d)  (2 j_d + 1) 
\nonumber \\
&\times &
 \sum_\kappa \left| R_{1, \kappa_d, \kappa}^{(E)} \right|^2 C\left(j_d\,  1 \, j ; \; \frac{1}{2}\,  0 \,  \frac{1}{2}\right)\, . 
\end{eqnarray}
Here, $\rho_i$ is the density of  initial continuum electronic states, $j$ and $j_d$  the total angular momentum of the electronic capture state and the continuum electron, respectively, and  $C(j_d\, 1\, j ; \; 1/2\ 0\  1/2)$ the Clebsch-Gordan coefficients as defined in Ref.~\cite{Edmonds}. The electric radial integral is given by
\begin{align}\label{radial_integral}
 R_{1, \kappa_d, \kappa}^{(E)} = \int_0^\infty \mathrm{d}r_e \left[ f_{n_d \kappa_d}(r_e) f_{\varepsilon \kappa} (r_e) + g_{n_d \kappa_d}(r_e) g_{\varepsilon \kappa} (r_e) \right] \, ,
\end{align}
with  $f_{\varepsilon \kappa} (r_e)$ and $g_{\varepsilon \kappa} (r_e)$ the continuum electron wave function components; $\varepsilon$ is the continuum electron energy measured from ionization threshold, defined by $\varepsilon = \sqrt{p^2c^2 + c^4} -c^2$. Furthermore, $f_{n_d \kappa_d}(r_e)$ and $g_{n_d \kappa_d}(r_e)$ are the respective components of the bound electron's wave function. As a further notation,  $n$ and $\kappa$  are the principal angular momentum  and the Dirac angular momentum quantum numbers, respectively, of the now bound electron (indexed $d$) and the free electron (no index), respectively.

\subsection{SMF}
The measure for the impact of NEEC on highly excited states for s-process nucleosynthesis is the  SMF defined in Eq.~\eqref{eq:smf2}. 
We are interested in this ratio for the cross section $ \sigma_{(n, \gamma)}$ for neutron capture followed by $\gamma$ decay, since it directly informs about the rate at which neutrons are permanently captured into the mother nucleus. 
Using the expressions for the relevant flux and cross sections, we obtain for  case I which does not take into account the possible effect of NEEC due to the electrons in the astrophysical plasma
\begin{equation}
  \langle \sigma v \rangle^\text{I} =  \int  \mathrm{d}E_n\phi(E_n) \frac{\pi\hbar^2}{p_n^2} \frac{\Gamma_{ex} \Gamma^\text{I}_{1}}{(E_{n} - E^\text{res}_{n})^2 + (\Gamma^\text{I}_{1}/2)^2}\frac{\Gamma^\text{I}_{\gamma_1}}{\Gamma^\text{I}_{1}}\, ,
\label{scenarioI}
\end{equation}
where $\Gamma^\text{I}_{1}$ is the total width of state (1) and $\Gamma^\text{I}_{\gamma_1}$ the $\gamma$ decay width corresponding to the decay rate $\lambda_\gamma^\text{I}$. The equivalent quantity for scenario II consists of two terms. A first term describes the process of neutron capture followed by $\gamma$ decay via state (1), while the second term stands for neutron capture, NEEC and $\gamma$ decay of state (2). 
Accordingly, we can write the rate $\langle \sigma v \rangle^\text{II}$ as
\begin{widetext}
\begin{eqnarray}
\langle \sigma v \rangle^\text{II} &=& \int  \mathrm{d}E_n\phi(E_n) \frac{\pi\hbar^2}{p_n^2}\frac{\Gamma_{ex} \Gamma^\text{II}_{1}}{(E_{n} - E^{res}_{n})^2 + (\Gamma^\text{II}_{1}/2)^2} 
\left[\frac{\Gamma^\text{II}_{\gamma_1}}{\Gamma^\text{II}_{1}} \right.
\nonumber \\
&&+\left.\frac{\hbar\Gamma^\text{II}_{\gamma_2}}{\Gamma^\text{II}_{1} \Gamma^\text{II}_{2}}\int  \mathrm{d}E_e
\mathrm{d}E_e^\text{res} \rho(E_{\alpha} + E^\text{res}_e-E_b)
\phi(E_e) \frac{\pi\hbar^2}{p_e^2} \frac{\Gamma_{\rm NEEC}\Gamma^\text{II}_{2}}{(E_{e} - E^\text{res}_{e})^2 + (\Gamma^\text{II}_{2}/2)^2} \right]\, .
\end{eqnarray}
\end{widetext}
Here, $\Gamma^\text{II}_{\gamma_2}$ and $\Gamma^\text{II}_{2}$ are the partial $\gamma$ and total widths of state (2), respectively. 
Furthermore, $\Gamma^\text{II}_{1}$ is the total width of state (1) in scenario II which  due to the (small) NEEC contribution will be slightly larger than $\Gamma^\text{I}_{1}$. As a consequence, the first (and leading) term  in the equation above will be smaller than the expression in Eq. (\ref{scenarioI}). Due to the different $\Gamma_{\gamma}/\Gamma$ branching ratios, this difference cannot be compensated by the second term  and the overall SMF values should be smaller than one. For NEEC occurring into excited electronic states, an additional branching ratio accounting for the decay of the captured electron to the atomic ground state should be taken into account. For practical cases however this factor will be one with a five-digit accuracy.

Both neutron and electron fluxes are taken into account by simultaneously integrating over  $E_n$ and $E_e$. 
The integration over continuum neutron energies $E_n$ can be performed stepwise for each neutron resonance individually, since the neutron resonances themselves are only several tens of $\mathrm{meV}$ wide and $\Gamma_{\rm NEEC}$ varies only insignificantly over the range of a single neutron resonance. In addition, the continuum nuclear level spectrum introduces an integration over the NEEC resonance energy $E_e^\text{res}$, see Eqs.~(\ref{sig_NEEC_single})-(\ref{eq:signa_final}).

\section{Results \label{results}}
In this section we present our numerical results for NEEC and photoabsorption rates in the stellar environment and for the SMF. The choice of studied isotopes relies on a scan of the nuclear chart for species for which the processes of interest have the most relevant contribution. The considered plasma conditions are typical for s-process nucleosynthesis sites. A calculation based on the Saha equation renders the probabilities for various charged states and ionization stages of the isotopes of interest, providing the input data for the NEEC atomic shell calculations. In the following subsections we detail these aspects and discuss our results.

\subsection{Test isotopes\label{sec:isotopes}}
As test cases for our analysis we chose $^{187}\mathrm{Os}$ and $^{193}\mathrm{Ir}$. The $^{187}\mathrm{Os}$ isotope is particularly relevant for astrophysics since it constitutes an important part of the Re/Os-clock. This ``clock'' is formed by a sequence of isotopes around the valley of stability whose singular decay properties allow the estimation of the galactic age. A further justification of our choice is related to our desire to investigate the isotopes for which a further excitation step of the compound nucleus is most probable and therefore of maximal relevance. 

Our model critically depends on nuclear level density and photon strength function parameters. Within the reduced transition probability estimation from the photon strength function, the final $(\text{NEEC},\delta)$ cross section depends  on $\rho$ and $S$ as: 
\begin{align} 
  \sigma_{\text{(NEEC}, \delta)}(E_e) & \propto \rho(E_\alpha + E_e-E_b)  B(E_\alpha \rightarrow E_\alpha + E_e-E_b)  \nonumber \\
  & \propto \rho(E_\alpha + E_e-E_b) \frac{S(E_e-E_b)}{\rho(E_\alpha)} \, .
   \label{eq:cross_section_dependency}
\end{align}
As $\rho(E_\alpha + E_e-E_b) \simeq \rho(E_\alpha)$ for relevant continuum and binding electron energies ($\sim 10-100\,  \mathrm{keV}$), cf. Fig.~\ref{psf}(b) , we find that the critical parameters are those determining the strength function. With this in mind we conducted a scan of the strength functions of all relevant s-process isotopes. As already noted at the end of Sec.~\ref{sec:Bees},   our estimate of the reduced transition probability heavily relies on the low energy tail of the photon strength function up to about $1\, \mathrm{MeV}$ (see also Fig.~\ref{psf}).
We have therefore compared the values for the integrated photon strength for energies up to $1\, \mathrm{MeV}$ for different isotopes.
Choosing a fixed benchmark for the energy, e.g. comparing the values of $S$ for $E_e -E_b= 0.1\, \mathrm{MeV}$ for different isotopes, gives equivalent results. 

\begin{figure}
  \centering
 \includegraphics[width=0.8\linewidth]{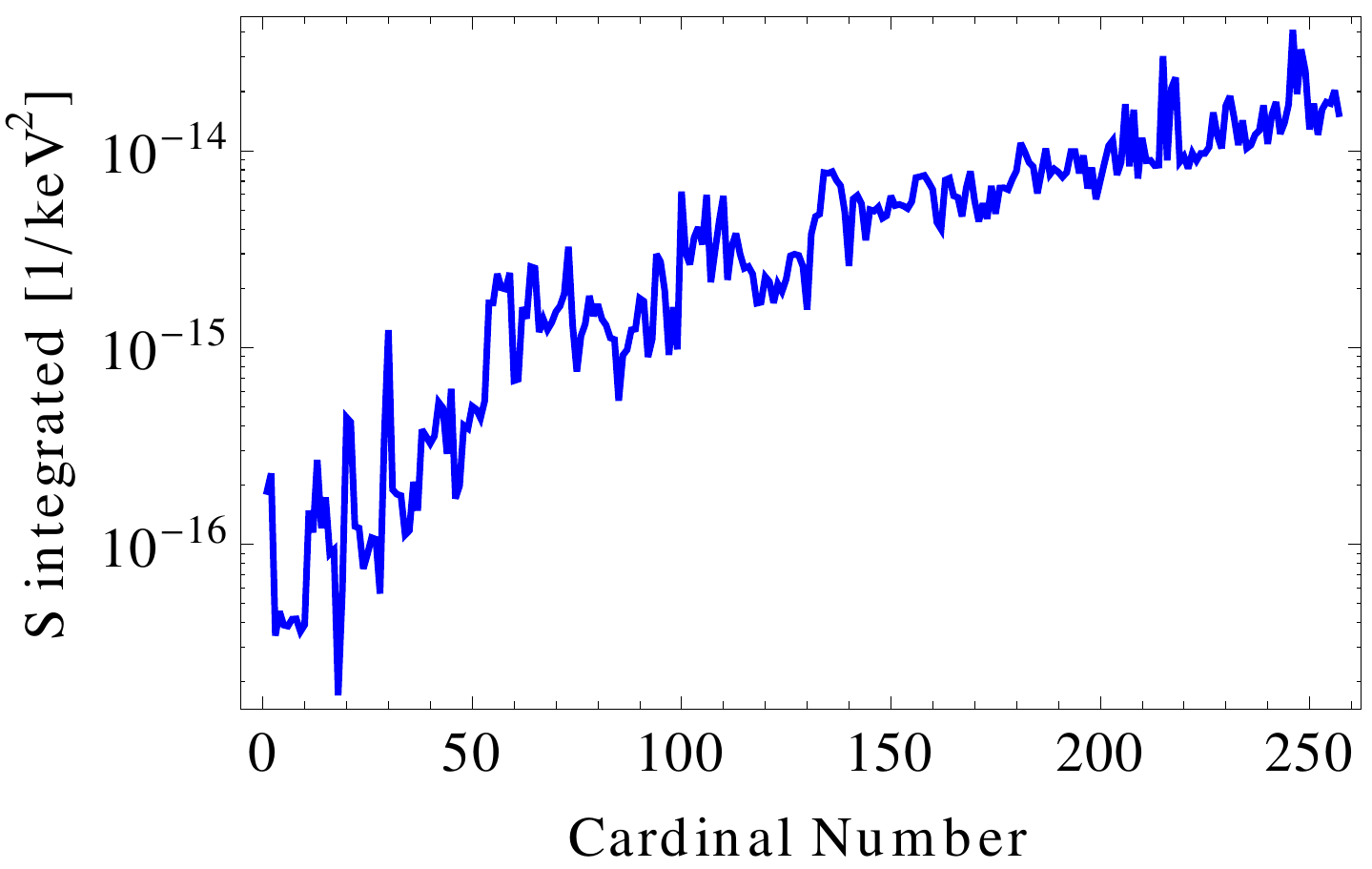}
  \caption{Integrated photon strength function up to 1 MeV for our isotope scan (based on the data tabulated in Ref.~\cite{Plujko2011567}). On the abscissa we use the cardinal number of the  parameter tables \cite{Plujko2011567}. Isotopes there are listed according to their mass number $A$ including also species which are not produced in the s-process.}
  \label{im:psf_search}
\end{figure}

\begin{figure}
  \centering
  \includegraphics[width=0.8\linewidth]{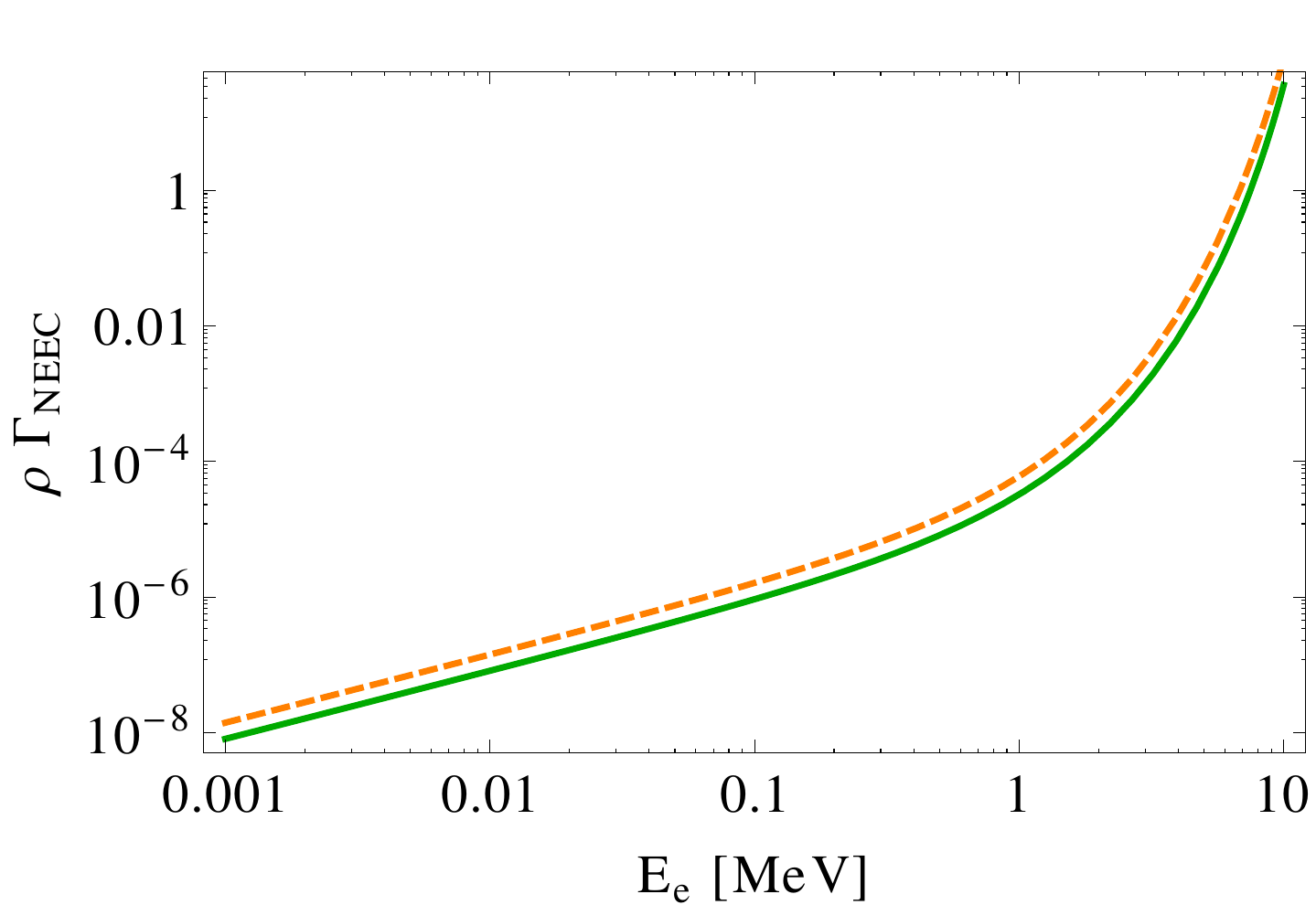}
  \caption{(color online). Comparison of the dependence of the $\sigma_{(\text{NEEC}, \delta)}$ cross section, i.e., the product between photon strength and nuclear level density as shown in Eq.~\eqref{eq:cross_section_dependency}, on the continuum electron energy for $^{187}\mathrm{Os}$ (solid green line) and $^{193}\mathrm{Ir}$ (dashed orange line).}

  \label{im:Os_Ir}
\end{figure}

Results for our scan based on parameters given in Ref.~\cite{Plujko2011567}  are presented in Fig.~\ref{im:psf_search}. The photon strength function was found to raise approximately linearly with cardinal number for isotopes belonging to the s-process chain. 
As a second candidate we chose $^{193}\mathrm{Ir}$  to be investigated as one of the isotopes with highest photon strength functions,  about two times larger  compared to the one of $^{187}\mathrm{Os}$. Fig.~\ref{im:Os_Ir} presents a comparison of the dependence of $\sigma_{(\text{NEEC}, \delta)}$ (i.e., the product of photon strength and nuclear level density) on the continuum electron energy as given in Eq.~\eqref{eq:cross_section_dependency} for both isotopes. The difference between the two isotopes turns out to be only marginal.  In the following we will therefore focus  solely on $^{187}\mathrm{Os}$ and only quote numerical results for $^{193}\mathrm{Ir}$ towards the end of the section.

\subsection{Plasma conditions \label{sec:plasma}}
The formation of $^{187}\mathrm{Os}$ (and $^{193}\mathrm{Ir}$) is at present ascribed to the strong main component of the s-process  assumed to take place in low-mass thermally pulsing asymptotic giant branch (TP-AGB) stars \cite{Son04}. Based on this scenario, the physical conditions that we adopt for our calculations refer to the $^{13}\mathrm{C}$ pocket site \cite{Ili07} in a low-mass TP-AGB star. Such carbon pockets consist mainly of helium and carbon, and to a minor extend also of oxygen. The few protons and the heavier elements mixed down from the convective envelope of the star in the third dredge-up \cite{GAB98} as well as the s-process products already formed are neglected since their abundance is barely noticeable compared to the major components. The s-process starts at temperatures $T\sim 0.9\cdot 10^8$ K, at which the lighter plasma constituents are already fully ionized. Assuming a 75$\%$ He and 25$\%$ C plasma composition, the number density of electrons in the plasma can be approximated as 
\begin{equation}\label{electron_number_density_mesured}
 n_e = 3 \frac{\rho}{3m_\text{He}/4+m_\text{C}/4}\, , 
\end{equation}
with $\rho$ being the plasma density.  We have disregarded here pair production, which is negligible at the temperature attained in the stellar interior---the rest mass of an electron is $m_e c^2 = 510.9$ keV, thus the photon energy threshold for pair production is reached only at approx.~1 MeV, while only temperatures of the order of 10--100 keV are of interest here.
The calculation of the electron flux requires knowledge of the chemical potential. For both electrons and neutrons, we obtain the chemical potentials using the normalization condition~\eqref{number_density}, which requires assumptions for the particle densities. For the electrons,  a reasonable choice for the plasma density at the temperatures of interest is~\cite{Takahashi1983}
\begin{equation}
 \rho \approx 10^3\, \mathrm{g\, cm^{-3}}\,.
\end{equation}
Although this value was estimated in Ref.~\cite{Takahashi1983}  for a pure He plasma, we find that the corresponding change in electron density for our composition of interest is not substantial. Based on these values, the electronic chemical potential is found to be $\approx -5 k_B T$.

For the case of neutrons, we can use directly measured data for the neutron particle density \cite{Kaeppler1989},
\begin{equation}\label{neutron_number_density_measured}
 n_n = 10^8\, \mathrm{g \,cm^{-3}}\, . 
\end{equation}
The neutron chemical potential is then $\approx -55 k_B T$, with $\mu_n/(k_BT)\ll 1$. In this case, quantum effects become negligible and 
a description for neutrons using the Maxwell-Boltzmann distribution is appropriate. This is also the  standard procedure with Maxwellian-averaged cross sections used in stellar neutron capture nucleosynthesis calculations~\cite{Beer1992}.
\begin{figure}[h!]
  \centering
  \includegraphics[width=0.8\linewidth]{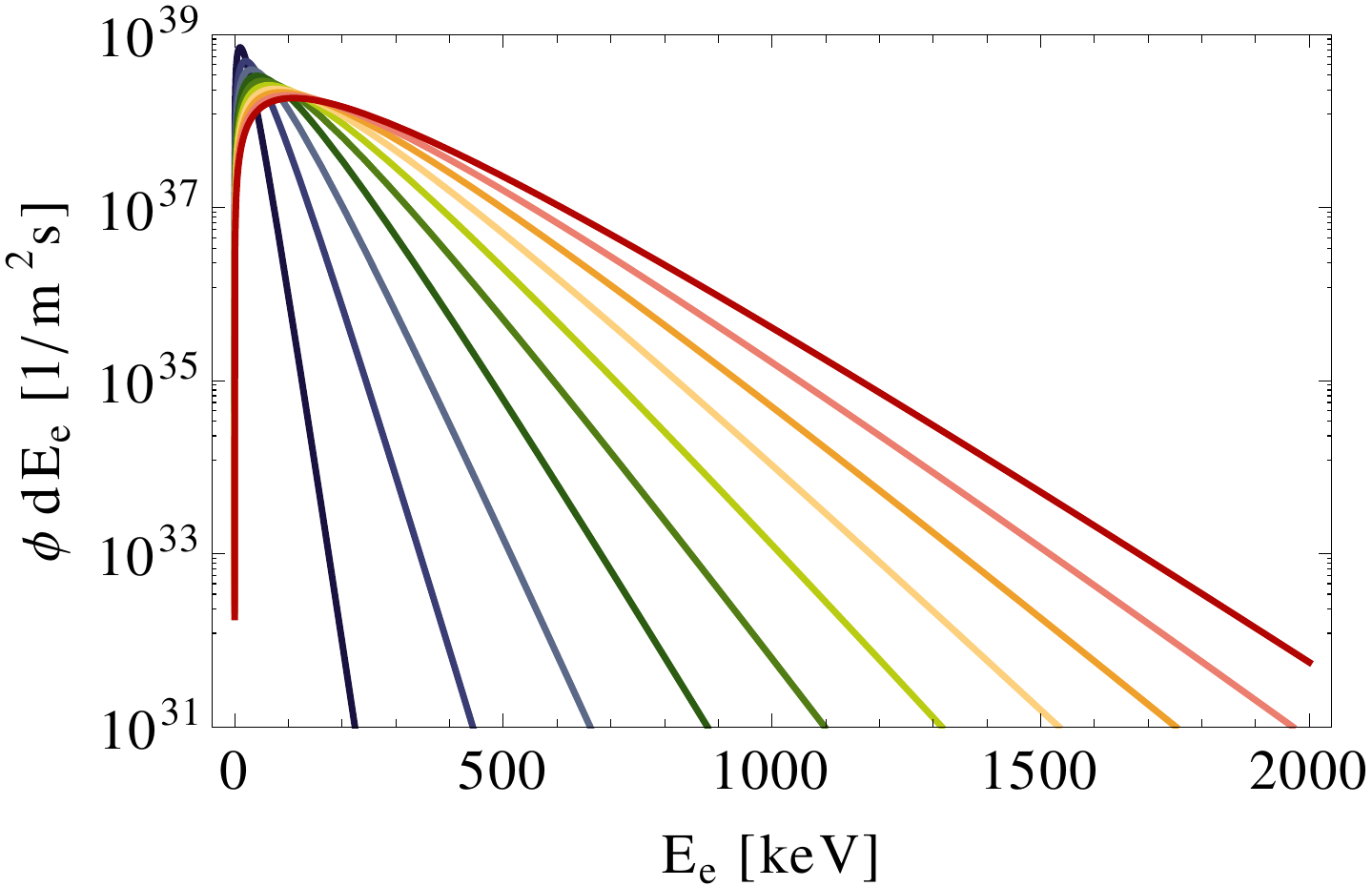}
  \caption{(color online). Differential electron flux for temperatures ranging from 10 to 100 keV. The lowest temperature (dark blue, most left) shows the steepest curve, with steepness decreasing with temperature for the other curves (increasingly warmer hue, from left to right). }
  \label{im:flux}
\end{figure}

We proceed to  investigate the differential energy flux~\eqref{flux}. The numerical integrations required for our calculations should be performed over a meaningfully chosen integration domain. The differential flux of continuum electrons depicted in Fig.~\ref{im:flux} reveals that even for highest temperatures under consideration any significant contribution to the rates will be suppressed by the decreasing flux above $E\sim2000\, \mathrm{keV}$. Furthermore, for both electrons and neutrons, the $1/p^2$ dependence of the cross sections of interest also favors capture of low-energy particles. The differential neutron flux is in addition heavily suppressed by the smaller chemical potential.  Hence we have chosen the integration bounds to coincide with the upper limit of available measured cross section data for the neutron reactions which are required for the calculation, i.e., few keV. For the numerical results the required nuclear cross section data was taken from Ref.~\cite{sigma_retrieval} and decay rates from Ref.~\cite{Mughabgab, n_tof_III}.

\subsection{Atomic charge states in the plasma \label{sec:vacancies}}
NEEC requires as prerequisite the existence of vacancies in the atomic shells. The charge state distribution in the plasma can be calculated  with the help of the Saha equation \cite{Saha20} assuming the ions in thermodynamical equilibrium. Starting from thermodynamical considerations, the Saha equation gives the ratio of two neighboring ionization stages for a certain atom. Since both the electronic chemical potential and the electron density are unknown, the Saha equation needs to be solved self-consistently \cite{Takahashi1983}. 

An important ingredient influencing the quantities in the Saha equation is the continuum depression in the plasma \cite{Mor82}. In order to estimate it we have followed the finite-temperature Thomas-Fermi model extended by Steward and Pyatt to include neighboring ions, valid for non-degenerate and non-relativistic electrons \cite{SPJ66}. This model was then extended to account for relativistic electrons and the quantum mechanical behavior of the bound electrons.

\begin{table}[h!]
\centering
\begin{tabular}{lcc}
  \hline
 & & \tabularnewline[-0.2cm]
Plasma composition    & $0.9\cdot 10^8$ K \hspace{0.2cm} &  $3.5\cdot 10^8$ K \\
 \hline
 & & \tabularnewline[-0.2cm]
78$\%$ He, 20$\%$ C, 2$\%$ O & 70.29 & 75.20\\
73$\%$ He, 25$\%$ C, 2$\%$ O & 70.29 & 75.20 \\
 75$\%$ H,  25$\%$ He & 69.31 &  74.90\\
100$\%$ He & 70.31 &  75.16 \\
 \hline
 \end{tabular}
 \caption{Average ionization state for the Os ion for two different plasma temperatures and four plasma compositions. } 
 \label{osmiumIon}
 \end{table}

We have performed calculations for four different plasma configurations and temperatures ranging from 0.9$\cdot 10^8$ K to 3.48$\cdot 10^8$ K. The resulting relative osmium ion number densities for T=0.9$\cdot 10^8$ K are distributed between charge states $66+$ and $74+$, whereas for T=3.48$\cdot 10^8$ K they range from $\mathrm{Os}^{71+}$ to $\mathrm{Os}^{76+}$. Our method was double-checked by comparing the calculated relative abundance of the different ionic charge states with results from Ref.~\cite{Takahashi1983} for the case of a pure He plasma. We find very good agreement between our values and the ones in \cite{Takahashi1983}, with deviations of less than 1$\%$. 

\begin{figure}[h!]
  \centering
  \includegraphics[width=0.8\linewidth]{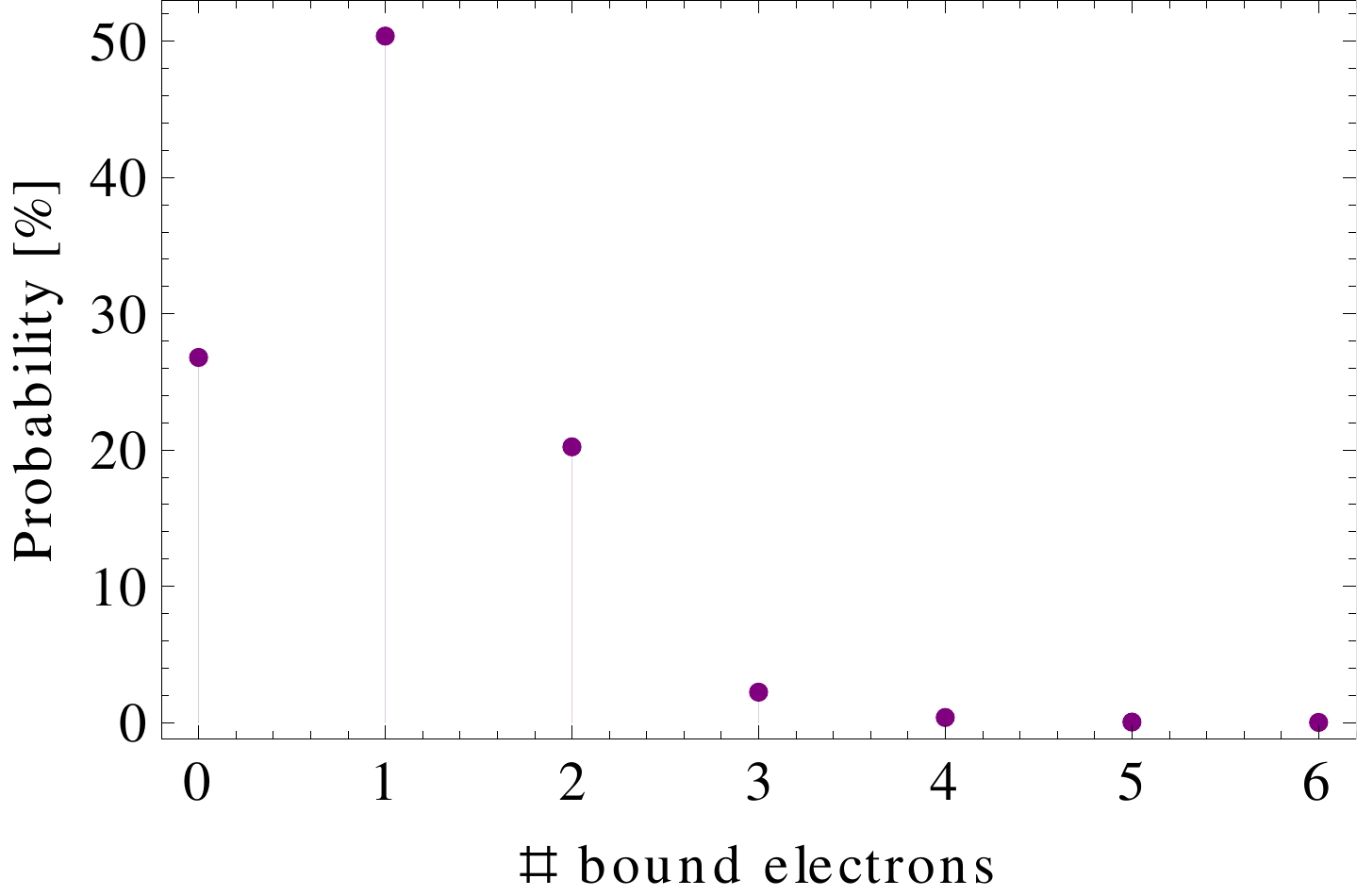}
  \caption{Ionization degree of $^{187}\mathrm{Os}$ at $T \approx 3\cdot 10^8$ K (corresponding to $k_BT$=30 keV) and $\rho \approx 10^3\,\mathrm{g\, cm^{-3}}$. The considered plasma composition is 78$\%$ He, 20$\%$ C, 2$\%$ O.}
  \label{im:ionization}
\end{figure}

The average osmium charge states in the plasma are given in Table \ref{osmiumIon}. We see that the differences between a pure He plasma composition and the He--C--O mixed composition characteristic of the C pocket nucleosynthesis site are very small.
An illustration of the relative ionic abundances as a function of the number of  bound electrons  left is presented in Fig.~\ref{im:ionization}. We conclude that in the stellar plasma, most osmium atomic shells are free. In particular, we are interested in NEEC into the $L$ shell, for which the cross sections are largest. Our results show that this atomic shell is most likely vacant and can serve as capture state for NEEC. The (on average) single bound electron is assumed to reside in the $1s$ orbital.

\subsection{Nuclear excitation rates and SMF}\label{sec:res_excitation}

Following neutron capture, the formed compound nucleus can undergo further (low-energy) excitations via electrons or photons in the plasma. Excitation via  NEEC can occur when free electrons in the plasma recombine into the highly charged Os ions. We consider here that the existing bound electron or electrons occupy the $K$ shell. Capture can then occur in all upper electronic shells, with probabilities for deeper bound levels being largest due to the $1/p^2$ dependence of the NEEC cross sections. We will therefore consider as standard the capture into the $2p$ orbitals, for which the NEEC rate is largest.  A compilation of results  for electron capture into all orbitals up to the $M$-shell with one initial $1s_{1/2}$ bound electron at a stellar temperature of $T = 30\, \mathrm{keV}$ are presented in Table ~\ref{tab:rates}.  Irrespective of the shell, capture into a $p$-orbital is much more likely than into a $s$ or $d$-state since the transition rates are highly parity dependent. 

The electronic part of the NEEC rates requires knowledge of both continuum and bound relativistic electronic wave functions. The continuum electron is assumed to be insensitive to the internal structure of the nucleus; relativistic Coulomb-Dirac wave functions are used \cite{Eichler}, taking the further approximation of a point-like nucleus. The final, captured bound electron wave functions are calculated for each case using the multiconfigurational Dirac-Fock  GRASP92 package \cite{Grasp92}. The plasma-induced shift for the bound electron energies has only a small impact for the considered NEEC rates. 

The nuclear coupling to the atomic shell via NEEC is compared with the effect of photoabsorption (PA), which can also have important contributions for nuclear excited level populations \cite{Gosselin04,Gosselin07}.  For PA on compound nuclei we follow the calculations outlined in Ref.~\cite{Bernstein}. The cross section $\sigma_\text{PA}$ for photoabsorption reads
\begin{align}
  \sigma_\text{PA}(E_\gamma) = 3 (\pi \hbar c)^2 E_\gamma S(E_\gamma)\, , 
\end{align}
where $E_\gamma$ denotes the energy of impinging photons. Both estimates for $\sigma_\text{NEEC}$ and $\sigma_\text{PA}$ rely on the same measured quantities (the photon strength function) and are therefore directly comparable. 

\begin{figure}
  \centering
\includegraphics[width=0.8\linewidth]{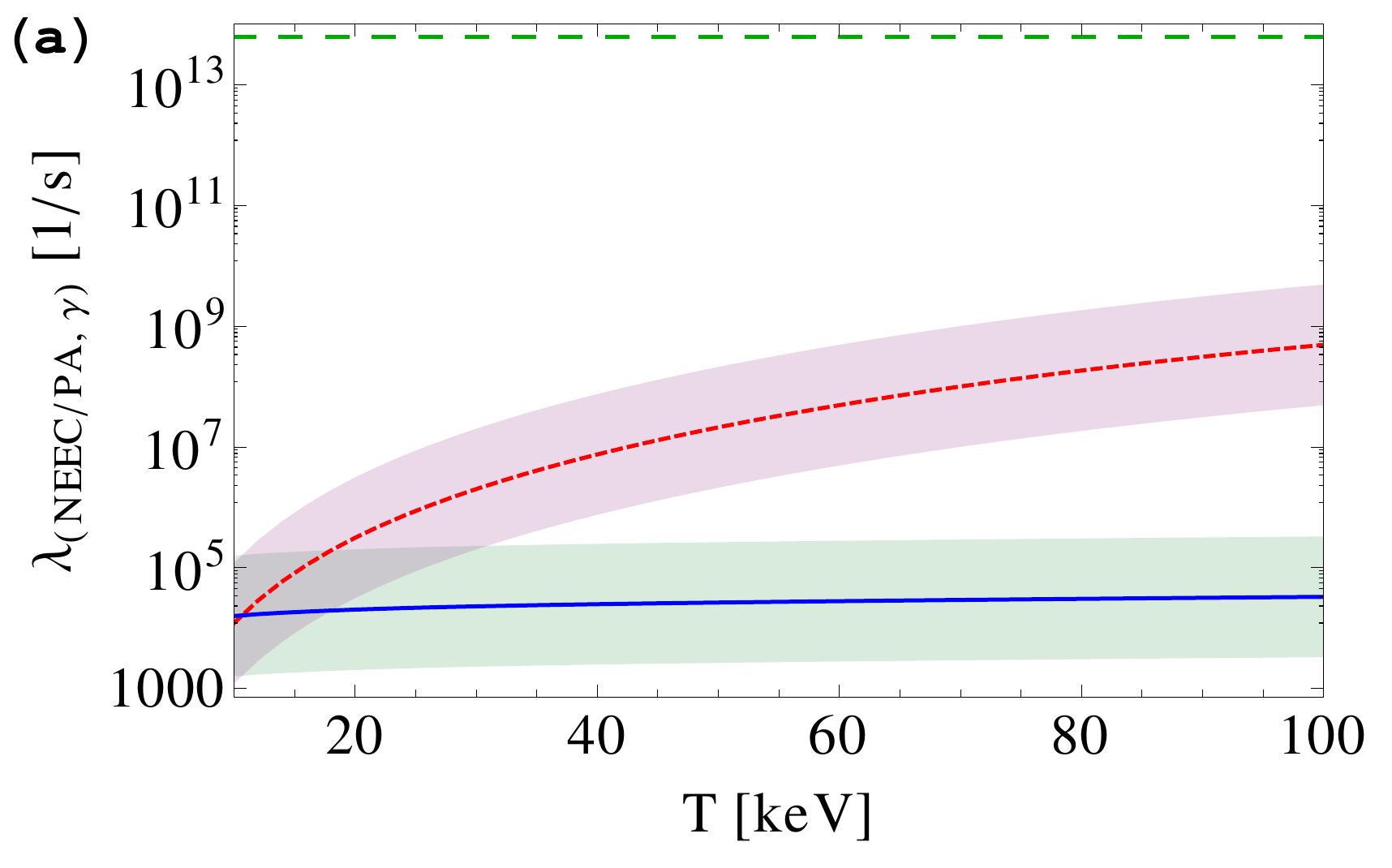}
\includegraphics[width=0.8\linewidth]{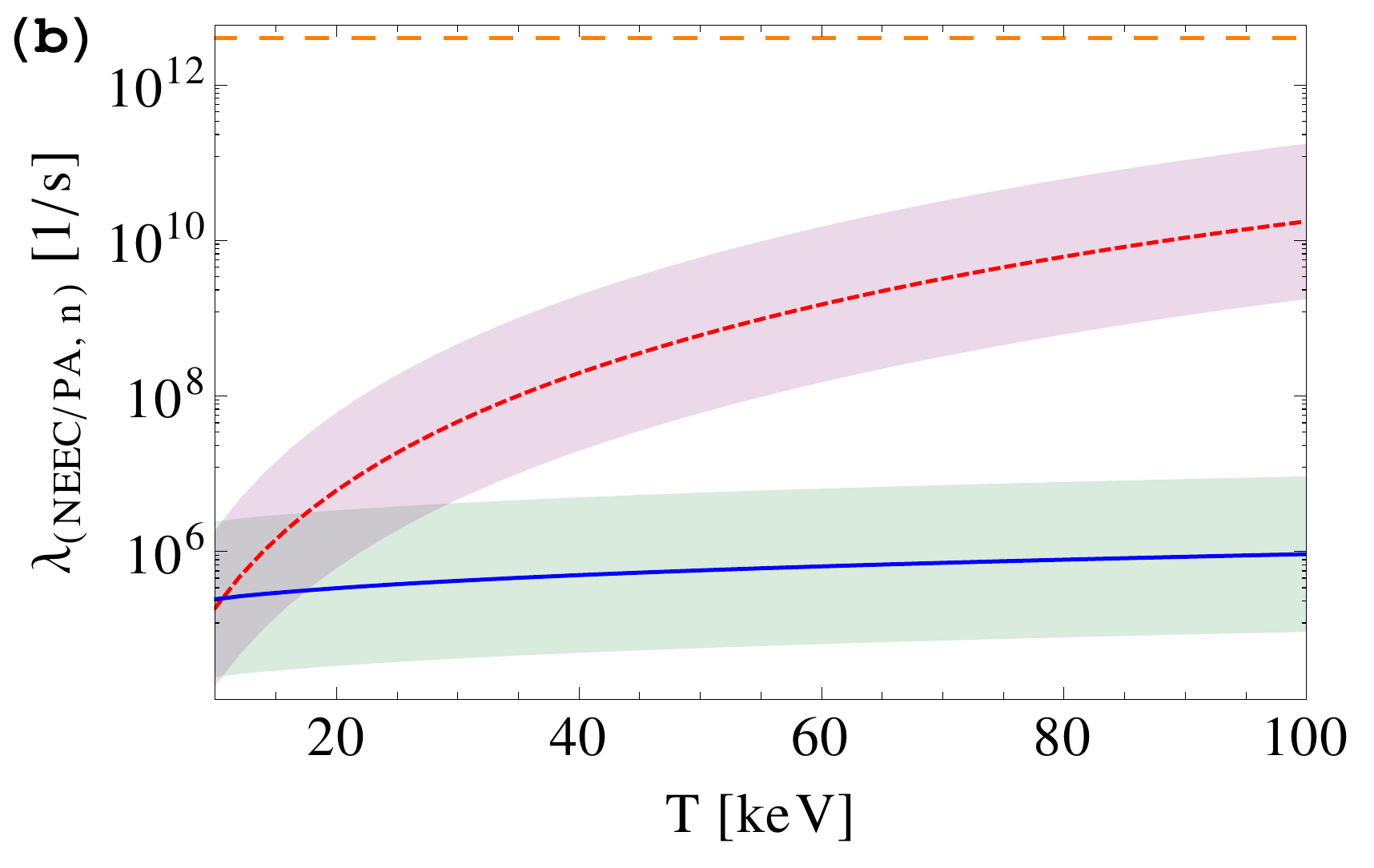}
  \caption{(color online). Reaction rates of NEEC into the $2p_{3/2}$ orbital (solid blue lines)  and PA (dashed red lines) followed by  (a) $\gamma$ decay or (b) neutron emission.  A one order of magnitude error spread due to the uncertainties in the extrapolation of the photon strength function at low energies was included by the shaded region for each estimate. The horizontal long-dashed lines represent the rates of the respective direct processes without including an additional NEEC or PA step. }
  \label{im:neecPlusDecay}
\end{figure}

We proceed to investigate the stellar rates for the excitation step built on the compound nucleus (NEEC or PA) followed by nuclear decay via neutron or $\gamma$ emission for the case of the energetically lowest neutron resonance.  Fig.~\ref{im:neecPlusDecay} shows the calculated rates for NEEC or PA, respectively, followed by either neutron re-emission or $\gamma$-decay.  To further illustrate and exemplify  this comparison we also present  the  respective PA rates in Table \ref{tab:rates} for a stellar temperature of $T = 30\, \mathrm{keV}$. While the $\lambda_{(\text{NEEC}, \delta)}$ rates vary only slowly with temperature, $\lambda_{(\text{PA}, \delta)}$ increases by several orders of magnitude within 10 to 100 keV temperature. 
However, both additional processes induced by NEEC and PA are less efficient by several orders of magnitude  compared to direct decay which is depicted by the horizontal long-dashed lines in Figs.~\ref{im:neecPlusDecay}(a,b).

\begin{figure}
  \centering
\includegraphics[width=0.8\linewidth]{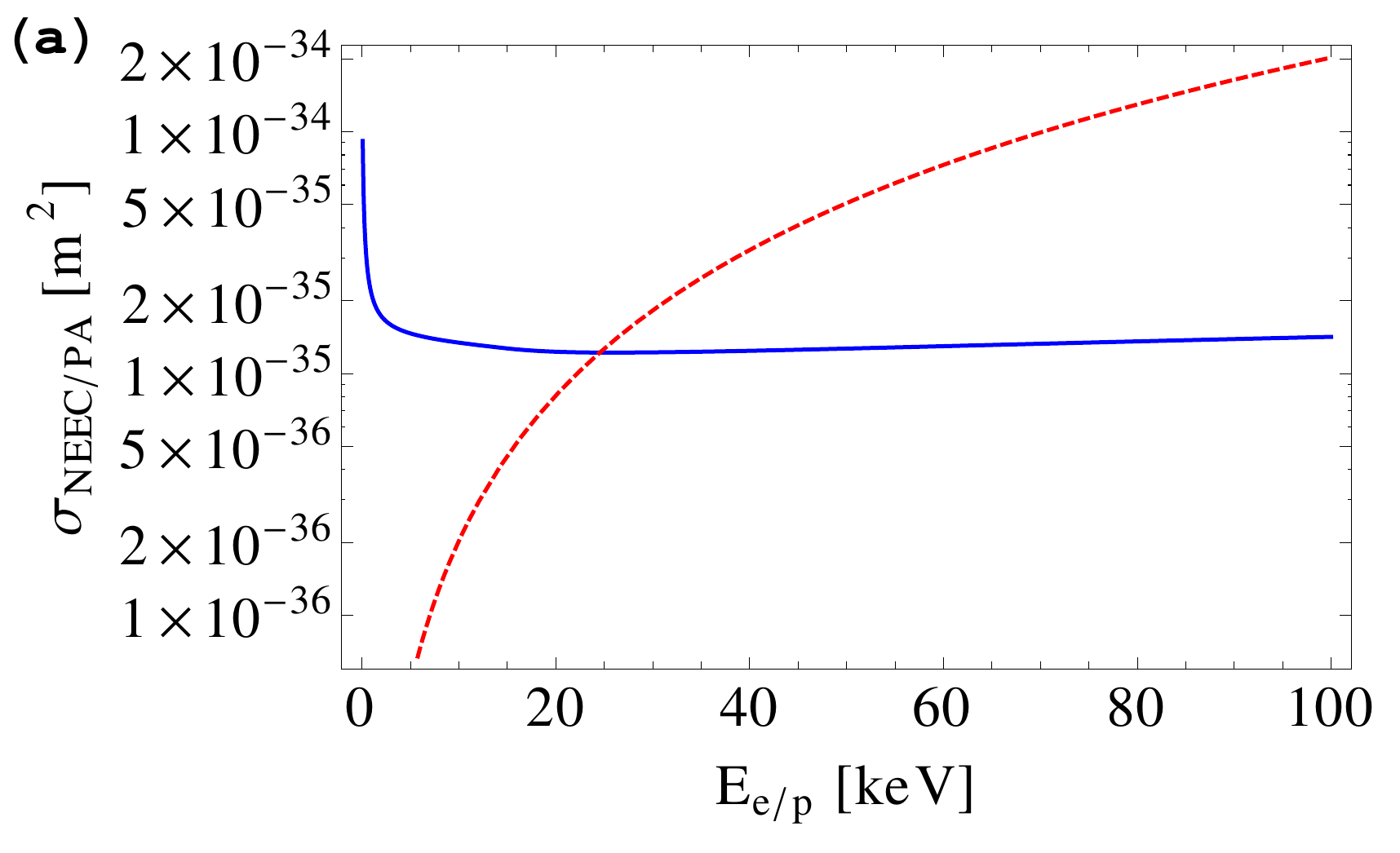} 
\includegraphics[width=0.8\linewidth]{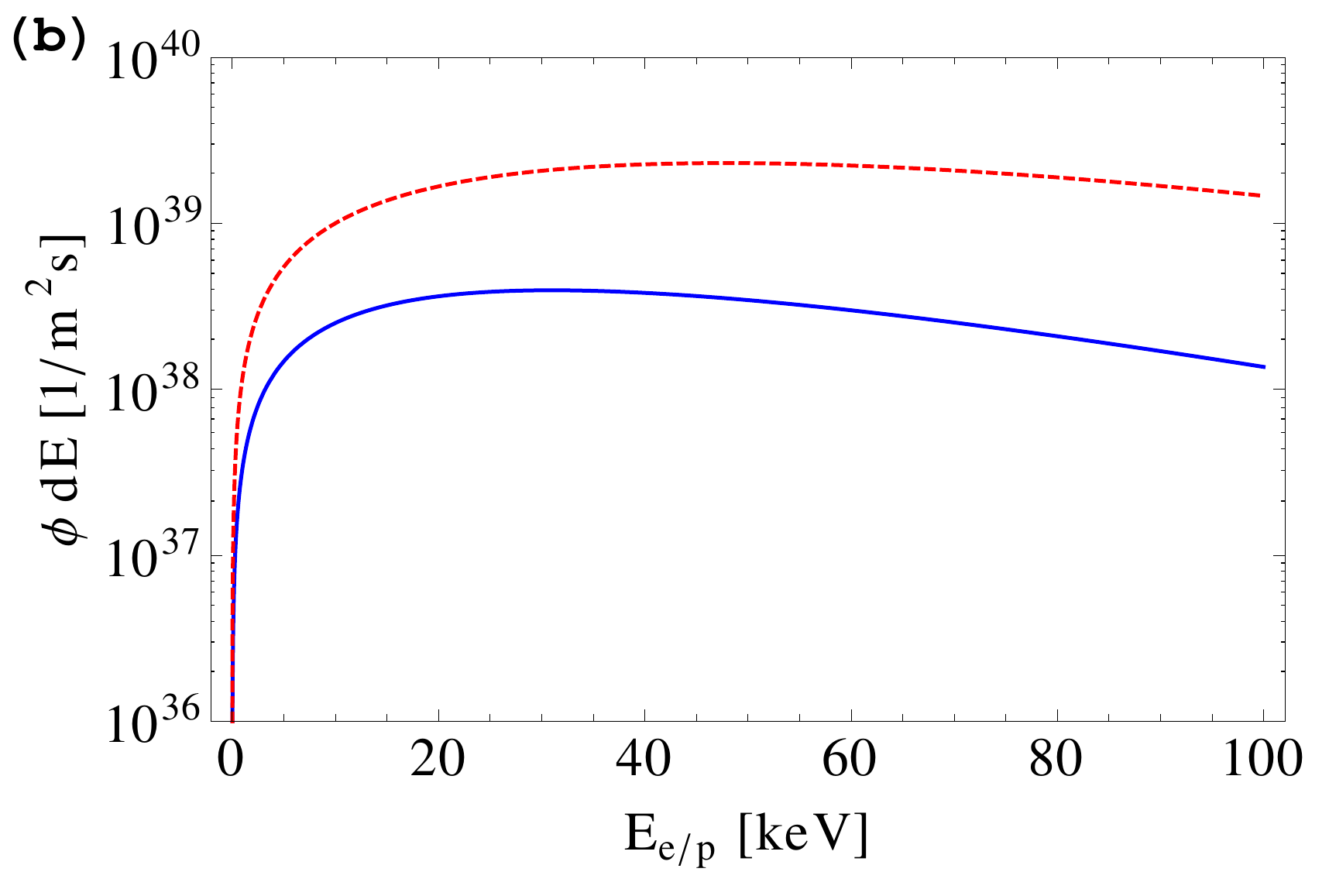}
  \caption{(color online) Comparison of cross sections (a) and fluxes (b) contributing in the  NEEC (solid blue line)  and PA (dashed red line) excitation rates as a function of excitation energy at $k_BT = 30\, \mathrm{keV}$.}
  \label{im:PhotEl}
\end{figure}

It is notable that the excitation rates for PA are well above the corresponding NEEC rates for all temperatures.  This may seem in contradiction with  previous results investigating the cross sections of the two processes  \cite{Pa07,PaJMO} and their corresponding plasma rates \cite{Gosselin04,Gosselin07,Gunst2014}. 
This behavior can be fathomed by investigating the flux and the cross section which are presented in Fig.~\ref{im:PhotEl}. Fig.~\ref{im:PhotEl}(b) shows that the photon flux is  over almost the entire  depicted energy range about one order of magnitude larger compared to the electron flux. This is due to the photons having zero chemical potential compared to the finite, negative one for electrons.  In contrast, the situation for the cross sections  is indeed according to the expected relation between the coupling to the atomic shell vs. coupling to the radiation field \cite{Pa07,PaJMO}. At low energies, the NEEC cross section is dominating and even diverging at $E = 0$ due to the $1/p^2(E)$ dependence in Eq.~\eqref{sig_general}. For higher energies however this suppresses the NEEC cross section compared to PA, which is then  substantially higher. The fluxes illustrated in Fig.~\ref{im:PhotEl}(b) indicate that relevant electron energies in the plasma far exceed the region where the NEEC cross section dominates over PA in the discussed scenario. As it was found in Ref.~\cite{Gunst2014}, lower temperatures on the order of hundreds of eV (in cold plasmas created by the impact of a x-ray free electron laser on metalic solid-state targets) indeed sufficiently reduce the relevant electron energy domain such that NEEC is rendered dominant.

\begin{table}
\centering
  \vspace{0.2cm}
\begin{tabular}{lcccc}

  \hline\\
 & & & & \tabularnewline[-0.5cm]
   & \multicolumn{2}{c}{$^{187}\mathrm{Os}$} & \multicolumn{2}{c}{$^{193}\mathrm{Ir}$}\\
  \cline{2-3} \cline{4-5} \tabularnewline[-0.3cm]
    & $\lambda_{\text{X} + \gamma}$ & $\lambda_{\text{X} + n}$ & $\lambda_{\text{X} + \gamma}$ & $\lambda_{\text{X} + n}$ \\
    $\text{X} = \text{NEEC}$ &&&& \\ 
  \hline & & \tabularnewline[-0.2cm]
  $1s_{1/2}$ & $1.69\times10^3$ & $4.35\times10^4$ & $3.81\times10^3$ & $7.84\times10^4$ \\
  $2s_{1/2}$ & $1.13\times10^3$ & $2.29\times10^4$ & $2.29\times10^3$ & $4.09\times10^4$ \\
  $2p_{1/2}$ & $1.16\times10^4$ & $2.14\times10^5$ & $2.33\times10^4$ & $3.83\times10^5$ \\
  $2p_{3/2}$ & $2.28\times10^4$ & $4.17\times10^5$ & $4.57\times10^4$ & $7.44\times10^5$ \\
  $3s_{1/2}$ & $9.92\times10^2$ & $1.94\times10^4$ & $2.01\times10^3$ & $3.44\times10^4$ \\
  $3p_{1/2}$ & $9.72\times10^3$ & $1.72\times10^5$ & $1.97\times10^4$ & $3.05\times10^5$ \\
  $3p_{3/2}$ & $1.93\times10^4$ & $3.41\times10^5$ & $3.91\times10^4$ & $6.05\times10^5$ \\
  $3d_{3/2}$ & $5.12\times10^2$ & $9.93\times10^3$ & $1.03\times10^3$ & $1.76\times10^4$ \\
  $3d_{5/2}$ & $1.49\times10^3$ & $2.93\times10^4$ & $2.67\times10^3$ & $5.24\times10^4$ \\
\hline & & \tabularnewline[-0.2cm]
  $\text{X} = \text{PA}$ 
  & $1.99\times10^6$ & $4.63\times10^7$ & $4.20\times10^6$ & $8.11\times10^7$\\
  \hline

\end{tabular}
\caption{Reaction rates for excitation  of the compound nucleus via NEEC or PA and subsequent decay at a stellar temperature of $k_BT = 30 \, \mathrm{keV}$. All rates are given in dimensions of  $\mathrm{[1/s]}$. The electron capturing orbital is given in the first row where appropriate. } 
\label{tab:rates}
\end{table}

\begin{figure}
  \centering
  \includegraphics[width=0.8\linewidth]{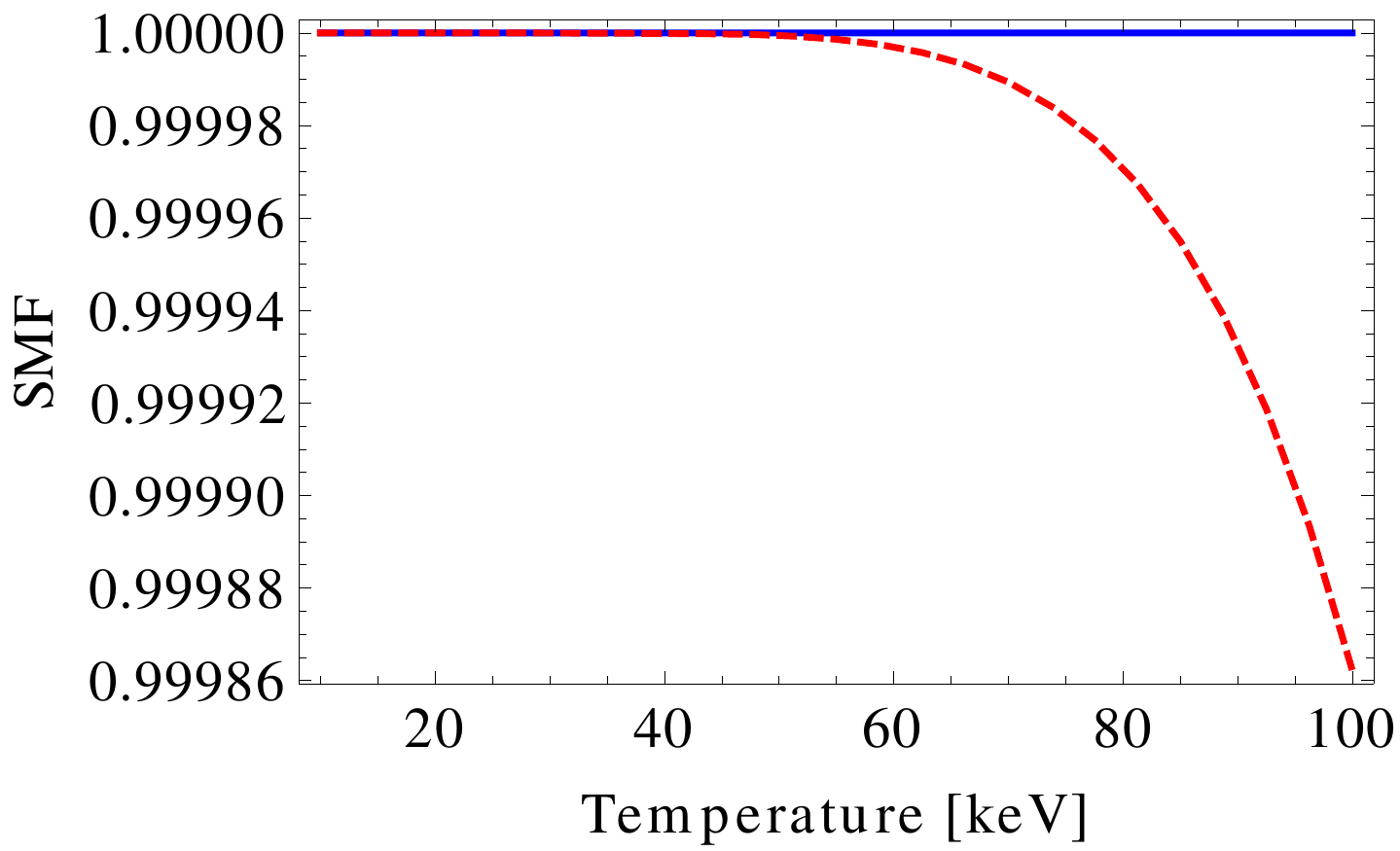}
  \caption{(color online). SMF as a function of the plasma temperature for the case of NEEC into the $2p_{3/2}$ orbital (dashed red line). The solid blue line illustrates SMF equal to unity.}
  \label{im:smf2}
\end{figure}

The results presented thus far suggest that both NEEC and PA of compound nuclei give only a small perturbation to the compound state's decay properties. The compilation of SMF values for $k_BT=30$ keV presented in Table \ref{tab:smf} further supports this assertion. For the considered temperature, the SMF differs from one by a very small amount of around $10^{-9}$ due to NEEC and $10^{-7}$ due to PA.  The SMF behavior with rising 
 temperatures considering the additional NEEC step is presented in Fig.~\ref{im:smf2}. We see here that the SMF deviance from unity is growing with rising temperature, reaching values of $10^{-4}-10^{-5}$ for $k_BT\simeq 100$ keV. In comparison, an equivalent PA-defined SMF is significantly decreasing with higher temperatures, reaching relevant values of $\simeq 0.1$ at the same temperature. The behavior of the NEEC and PA cross sections in Fig.~\ref{im:PhotEl} justifies this feature for PA. As the photon energies shift towards higher values,   greater energy transfer to the nucleus is rendered possible.  Our results support therefore the conclusions of Ref.~\cite{Bernstein} for the r-process nucleosynthesis scenario which typically takes place at  higher temperatures than the ones considered here.

\begin{table}[t]
\centering
\small
\begin{tabular}{lll}

  \hline \\
& & \tabularnewline[-0.5cm]
   & $^{187}\mathrm{Os}$ & $^{193}\mathrm{Ir}$\\
   NEEC & & \\
  \hline \tabularnewline[-0.3cm]
  $1s_{1/2}$ & $3\times10^{-10}$ & $5\times10^{-10}$ \\
  $2s_{1/2}$ & $2\times10^{-10}$ & $2\times10^{-10}$ \\
  $2p_{1/2}$ & $1\times10^{-9}$ & $2\times10^{-9}$ \\
  $2p_{3/2}$ & $3\times10^{-9}$ & $4\times10^{-9}$ \\
  $3s_{1/2}$ & $1\times10^{-10}$ & $2\times10^{-10}$ \\
  $3p_{1/2}$ & $1\times10^{-9}$ & $2\times10^{-9}$ \\
  $3p_{3/2}$ & $2\times10^{-9}$ & $4\times10^{-9}$ \\
  $3d_{3/2}$ & $7\times10^{-11}$ & $1\times10^{-10}$ \\
  $3d_{5/2}$ & $2\times10^{-10}$ & $3\times10^{-10}$ \\ 
  \hline \tabularnewline[-0.3cm]
  PA & $3\times10^{-7}$ & $5\times10^{-7}$ \\
  \hline

\end{tabular}
\caption{ Values of the variation of SMF from unity (1-SMF) at $k_BT = 30\, \mathrm{keV}$ for different NEEC capture atomic orbitals and for PA. } 
\label{tab:smf}
\end{table}

\section{Conclusions \label{conclusion}}
We have investigated the impact of a further low-energy excitation step via NEEC or PA on the decay channels of a compound nucleus formed by neutron capture in astrophysical s-process sites. To this end we have introduced and calculated the SMF as a highly excited nucleus-counterpart of the well-known SEF. Our results show that from the two considered processes, only PA can potentially switch the role of $\gamma$-decay and neutron reemission decay channels for the compound nucleus, provided that high plasma temperatures on the order of $k_BT=100$ keV are available. This is rather the parameter regime of the r-process nucleosynthesis. Due to its decreasing cross sections for high-energy electrons, NEEC starting from the compound nucleus is not competitive with the main decay channels $\gamma$ decay and neutron reemission and does not have a significant effect on the decay branching ratio relevant for nucleosynthesis. This holds true both at the lower temperatures dominating the s-process sites as well as the higher temperatures typical for the r-process. We conclude that the role of the coupling of the nucleus to the atomic shells in dense astrophysical plasmas is restricted to low-lying levels in the vicinity of the ground state or of specific isomeric states.


\bibliographystyle{apsrev}
\bibliography{biblio}

\end{document}